\documentclass{emulateapj}
\usepackage{natbib}
\usepackage{epsfig}
\usepackage{graphicx}
\usepackage{subfigure}
\usepackage{float}
\usepackage{amsmath}
\usepackage{color}
\usepackage{amssymb}
\usepackage{amsfonts}
\usepackage{apjfonts}
\usepackage[colorlinks,linkcolor=blue,anchorcolor=green,citecolor=blue]{hyperref}
\shorttitle{Model-independent Determination of Curvature Parameter}
\begin{document}

\title{Model-independent Curvature Determination from Gravitational-Wave Standard Sirens and Cosmic Chronometers}

\author{Jun-Jie Wei\altaffilmark{1,2}}
\affil{$^1$ Purple Mountain Observatory, Chinese Academy of Sciences, Nanjing 210008, China; jjwei@pmo.ac.cn\\
$^2$ Guangxi Key Laboratory for Relativistic Astrophysics, Guangxi University, Nanning 530004, China}

\begin{abstract}
The detection of gravitational waves (GWs) provides a direct way to measure the luminosity distance,
which enables us to probe cosmology. In this paper, we continue to expand the application of
GW standard sirens in cosmology, and propose that the spatial curvature can be estimated in a
model-independent way by comparing the distances from future GW sources and current cosmic-chronometer
observations. We expect an electromagnetic counterpart of the GW event to give the source redshift,
and simulate hundreds of GW data from the coalescence of double neutron stars and black hole--neutron star
binaries using the Einstein Telescope as reference. Our simulations show that, from 100 simulated GW events
and 31 current cosmic-chronometer measurements, the error of the curvature parameter $\Omega_{K}$ is expected
to be constrained at the level of $\sim0.125$. If 1000 GW events are observed, the uncertainty of
$\Omega_{K}$ would be further reduced to $\sim0.040$. We also find that adding 50 mock $H(z)$ data
(consisting of 81 cosmic-chronometer data and 1000 simulated GW events) could result in much tighter constraint on
the zero cosmic curvature, for which, $\Omega_{K}=-0.002\pm0.028$. Compared to some actual model-independent
curvature tests involving the distances from other cosmic probes, this method with GW data
achieves constraints with much higher precision.
\end{abstract}


\section{Introduction}

The spatial curvature of the universe is one of the important research topics in modern cosmology.
To be specific, estimating the cosmic curvature is an effective way to test the fundamental assumption
that the universe is well described by the homogeneous and isotropic Friedmann--Lema\^{\i}tre--Robertson--Walker
(FLRW) metric. Note that the possible invalidation of the FLRW approximation has been suggested to
explain the accelerated expansion of the universe
(e.g.,~\citealt{2006JHEP...02..016F,2008GReGr..40..451E,2009JHEP...04..006F,2009JCAP...02..011R,2013JCAP...09..003B,2013JCAP...12..051L,2014A&A...570A..63R}).
On the other hand, even if the FLRW metric is valid, whether the cosmic space is open, flat, or closed
is crucial for us to understand the evolution of our universe and the nature of dark energy
\citep{2006PhRvD..73h3526I,2006JCAP...12..005I,2007JCAP...08..011C,2007PhRvD..75d3520G,2008JCAP...12..008V,2015PhRvD..91h3536Z}.
Any significant deviation from the zero cosmic curvature
would have far-reaching consequences for fundamental physics and inflation models
\citep{2005ApJ...633..560E,2006PhRvD..74l3507T,2007ApJ...664..633W,2007PhLB..648....8Z}.
Although a spatially flat Universe ($\Omega_{K}=0$) is strongly supported by various
cosmological probes, especially by the latest \emph{Planck}2015 results of observations of the cosmic
microwave background (CMB;~\citealt{2016A&A...594A..13P}),\footnote{Using non-flat inflation model energy
density inhomogeneity power spectra \citep{1982Natur.295..304G,1984NuPhB.239..257H,1985PhRvD..31.1931R,1994ApJ...432L...5R,1995PhRvD..52.1837R,2017PhRvD..96j3534R},
some studies have found that the \emph{Planck}2015 CMB anisotropy data favor a mildly closed universe
\citep{2017arXiv170703452O,2017arXiv171003271O,2017arXiv171208617O,2018arXiv180100213P,2018arXiv180305522P}.
Besides, currently available non-CMB data do not significantly require zero spatial curvature
\citep{2015Ap&SS.357...11F,2017ApJ...835...26F,2016ApJ...829...61C,2017arXiv171200018M,2018arXiv180506408R}.}
most of the curvature constraints
are not in a direct geometric way. That is, some specific cosmological models (e.g., the standard
$\Lambda$CDM model) are assumed in a determination of the curvature, thus these results are indirect
and cosmological-model-dependent. Besides, because of the strong degeneracy between the cosmic
curvature $\Omega_{K}$ and the dark energy equation of state $w$, it is difficult to constrain
the two parameters simultaneously in a non-flat $w$CDM model. Therefore, it would be better to
measure spatial curvature by purely geometrical and model-independent methods.

In \citet{2006ApJ...637..598B}, a model-independent determination of the curvature parameter $\Omega_{K}$
based on the sum rule of distances along null geodesics of the FLRW metric was presented (see also \citealt{2006PhRvD..73b3503K}).
Recently, this distance sum rule was put forward to test the FLRW metric and estimate
the curvature by using the Union2.1 compilation of type Ia supernovae (SNe Ia) and
strong gravitational lensing systems observed from the Sloan Lens ACS Survey \citep{2015PhRvL.115j1301R}.
However, the spatial curvature was weakly constrained due to the large uncertainties in the
gravitational lensing data. Following the method of \citet{2015PhRvL.115j1301R}, the null test
of the curvature has been carried out with updated observations
\citep{2017ApJ...839...70L,2017ApJ...834...75X,2018ApJ...854..146L,2018arXiv180301990Q}.
Based on this distance sum rule, \citet{2018JCAP...03..041D} recently employed strong lensing time delays
and supernova distances to measure the curvature, and estimated uncertainties on the curvature
enabled by future survey data.
Another model-independent method was proposed to determine $\Omega_{K}$ by combining measurements
of the Hubble parameter $H(z)$ and the luminosity distance $D_{L}(z)$ (or the angular diameter distance $D_{A}(z)$)
\citep{2007JCAP...08..011C,2008PhRvL.101a1301C}:
\begin{equation}
\Omega_{K}=\frac{\left[H(z)D'(z)\right]^{2}-1}{\left[H_{0}D(z)\right]^{2}}\;,
\label{eq:clarkson}
\end{equation}
where $H_{0}$ is the Hubble constant, $D(z)=(1+z)D_{A}(z)=D_{L}(z)/(1+z)$ represents the comoving
angular diameter distance, and $D'(z)=dD(z)/dz$ denotes the derivative with respect to
redshift $z$. This method have been extensively used in the literature
\citep{2010PhRvD..81h3537S,2011arXiv1102.4485M,2014PhRvD..90b3012S,2014ApJ...789L..15L,2014PhRvD..89b3503Y,2016PhRvD..93d3517C,2017JCAP...01..015L,2017JCAP...03..028R}.
However, in this method, the derivative of comoving distance with respect to $z$ is necessary
to estimate the curvature, which introduces a considerable uncertainty. Recently, stricter
constraints on the curvature from measurements of expansion rate and distance have been
obtained by dodging the derivative of distance with respect to $z$
\citep{2016ApJ...833..240L,2016ApJ...828...85Y,2017arXiv170808608C,2017ApJ...847...45W,2017ApJ...838..160W,2018ApJ...856....3Y}.
{In addition to these above two methods, there has been some other works proposing model-independent methods
to determine the curvature and some approaches to reduce the measurement sensitivity
of $\Omega_{K}$ to dark energy (e.g., \citealt{2015PhRvD..92l3518T,2018MNRAS.477L.122W}).
For example, \citet{2015PhRvD..92l3518T} estimated a best achievable accuracy of the curvature constraint
with the radial and angular diameter distances from future baryon acoustic oscillation experiments.
\citet{2018MNRAS.477L.122W} showed that forthcoming 21 cm intensity mapping experiments are ideally designed
to carry out curvature determinations, as they can detect the clustering signal at high redshift
with sufficient precision to break the degeneracy of dark energy and curvature.

On the other hand, the joint detection of the gravitational-wave (GW) event GW170817 with electromagnetic (EM)
counterparts (e.g., a gamma-ray burst GRB 170817A, or a kilonova) from the merger of binary neutron stars (NSs)
\citep{2017PhRvL.119p1101A,2017Sci...358.1556C,2017ApJ...848L..14G} has opened the new era of multi-messenger
cosmology \citep{2017Natur.551...85A}. The application of GWs in cosmology was first suggested
by \citet{1986Natur.323..310S}, who proposed that the Hubble constant can be determined from GW observations,
since the waveform signals of GWs from inspiralling and merging compact binaries encode distance information
(see also~\citealt{2005ApJ...629...15H,2014PhRvX...4d1004M,2018PhRvD..97f4031Z}).
Thus, GWs can serve as standard sirens, analogous to SN standard candles. But,
unlike the distance calibration of SNe Ia that rely on the nuisance parameters characterizing SN light-curves,
the GW standard-siren observations can measure the luminosity distances directly, without the need of any other
cosmic distance ladders (i.e., they are self-calibrating). This advantage of GWs can help us dodge the influence
of the nuisance parameters on the test of the curvature, which should be considered when one makes use of
the SNe Ia data \citep{2016ApJ...833..240L,2017ApJ...838..160W}. Therefore, combining $H(z)$ and GW data
provides a novel way to determine the cosmic curvature.

In the past, the simulated GW data have been used to measure the cosmological parameters (e.g.,
~\citealt{2005ApJ...629...15H,2011PhRvD..83b3005Z,2012PhRvD..86d3011D,2016PhRvD..93d3517C,2017PhRvD..95d3502D,2017NatCo...8.1148L,2017MNRAS.472.2906W,2018arXiv180512265W}),
test the cosmic distance duality relation \citep{2017arXiv171010929Y}, weigh the total neutrino mass \citep{2018arXiv180204720W},
explore the anisotropy of the universe \citep{2018PhRvD..97j3005C,2018EPJC...78..356L},
and constrain the time variation of Newton's constant $G$ \citep{2018arXiv180403066Z}.
We note that one recent work \citep{2018JCAP...04..002J} provided an analysis of curvature constraints
in a model-independent way using distance probes: GWs, cosmic chronometers, and redshift drift.
They discussed what kind of observations and what level of uncertainty will be needed to measure
the curvature at the CMB fluctuations level of $\sim10^{-5}$, and found that one could measure the curvature
at the desired accuracy only when improving the uncertainties on the Hubble parameter and the luminosity distance
from the GW source by a factor of 10 and 1000, respectively.
In this paper, following the method proposed in \citet{2007JCAP...08..011C,2008PhRvL.101a1301C},
we investigate what level of curvature constraints can be achieved using future GW data in the era of the third-generation
GW detectors such as the Einstein Telescope (ET). The uncertainty on the GW luminosity distance
is adopted as the designed level of the ET.
Firstly, using a non-parametric smoothing technique,
we reconstruct a continuous $H(z)$ function from measurements of the expansion rate from cosmic chronometers.
The model-independent comoving distance can then be directly obtained by integrating the reconstructed $H(z)$ function.
Next, with the curvature parameter $\Omega_{K}$ taken into account, we transform the comoving distance into
the curvature-dependent luminosity distance $D_{L}^{H}$. Finally, by comparing $D_{L}^{H}(z)$ to
the luminosity distances $D_{L}^{\rm GW}(z)$ derived from the mock GW data, we achieve a cosmology-independent
and compelling test of the cosmic curvature.

The paper is arranged as follows. In Section~\ref{sec:method}, we derive the luminosity distance information $D_{L}^{H}$
and $D_{L}^{\rm GW}$ from expansion rate measurements and GW standard sirens, respectively.
In Section~\ref{sec:simulation}, we demonstrate that an accurate determination of the curvature parameter
can be achieved in a model-independent way by confronting $D_{L}^{H}$ with $D_{L}^{\rm GW}$, using Monte Carlo
simulations. Lastly, we give a brief summary and discussion in Section~\ref{sec:summary}.
Throughout this paper, the geometric unit $G=c=1$ is adopted.

\section{Method Description}
\label{sec:method}

\subsection{Distance from Cosmic-Chronometer Measurements}
Since the expansion rate of the universe relates to the expansion factor $a(t)$,
i.e., $H(z)\equiv\dot{a}/a$, $H(z)$ can be directly measured from the time-redshift derivative
$dt/dz$ using $H(z)=-\frac{1}{1+z}\frac{dz}{dt}$. That is, the Hubble parameter $H(z)$
can be obtained in a cosmology-independent way by calculating the differential age evolution
of passively evolving galaxies \citep{2002ApJ...573...37J}. In the literature, these galaxies
are usually called cosmic chronometers. We use the most complete sample of 30 cosmic-chronometer
measurements that obtained from \citet{2016JCAP...05..014M}. We also include a recent
cosmic-chronometer measurement at a redshift of $z=0.47$ \citep{2017MNRAS.467.3239R}.
Our sample now contains 31 data points in the redshift range of $0< z < 2.0$, which is listed in Table~\ref{table1}.

\begin{table}
\centering \caption{$H(z)$ Measurements Obtained from the Cosmic-Chronometer Approach}
\begin{tabular}{lcc}
\hline
\hline
 \emph{z} & $H(z)$ (km $\rm s^{-1}$ $\rm Mpc^{-1}$) & References \\
\hline
0.09	&	$	69	\pm	12	$	&	\citet{2003ApJ...593..622J} \\
\hline
0.17	&	$	83	\pm	8	$	&	\\
0.27	&	$	77	\pm	14	$	&	\\
0.4	&	$	95	\pm	17	$	&	 \\
0.9	&	$	117	\pm	23	$	&	\citet{2005PhRvD..71l3001S} \\
1.3	&	$	168	\pm	17	$	&	\\
1.43	&	$	177	\pm	18	$	&	\\
1.53	&	$	140	\pm	14	$	&	\\
1.75	&	$	202	\pm	40	$	&	\\
\hline
0.48	&	$	97	\pm	62	$	&	\citet{2010JCAP...02..008S} \\
0.88	&	$	90	\pm	40	$	&	\\
\hline
0.1791	&	$	75	\pm	4	$	&	\\
0.1993	&	$	75	\pm	5	$	&	\\
0.3519	&	$	83	\pm	14	$	&	\\
0.5929	&	$	104	\pm	13	$	&	\citet{2012JCAP...07..053M} \\
0.6797	&	$	92	\pm	8	$	&	\\
0.7812	&	$	105	\pm	12	$	&	\\
0.8754	&	$	125	\pm	17	$	&	\\
1.037	&	$	154	\pm	20	$	&	\\
\hline
0.07	&	$	69	\pm	19.6	$	&	\\
0.12	&	$	68.6	\pm	26.2	$	&	\citet{2014RAA....14.1221Z} \\
0.2	&	$	72.9	\pm	29.6	$	&	\\
0.28	&	$	88.8	\pm	36.6	$	&	\\
\hline
1.363	&	$	160	\pm	33.6	$	&	\citet{2015MNRAS.450L..16M} \\
1.965	&	$	186.5	\pm	50.4	$	&	\\
\hline
0.3802	&	$	83	\pm	13.5	$	&	\\
0.4004	&	$	77	\pm	10.2	$	&	\\
0.4247	&	$	87.1	\pm	11.2	$	&	\citet{2016JCAP...05..014M} \\
0.4497	&	$	92.8	\pm	12.9	$	&	\\
0.4783	&	$	80.9	\pm	9	$	&	\\
\hline
0.47	&	$	89	\pm	50	$	&	\citet{2017MNRAS.467.3239R} \\
\hline
\end{tabular}
\label{table1}
\end{table}

In our analysis, we use the model-independent smoothing technique, Gaussian process (GP),
to reconstruct a continuous $H(z)$ function that best approximates the discrete Hubble
parameter data we have compiled in Table~\ref{table1}. There is an open-source Python package
of GP called GaPP developed by \citet{2012JCAP...06..036S}, which is widely used for cosmological studies
(e.g.,~\citealt{2012MNRAS.425.1664B,2012PhRvD..86h3001S,2016PhRvD..93d3517C,2016ApJ...828...85Y,2017ApJ...838..160W,2017JCAP...11..029Y,2018EPJC...78..258Y,2018JCAP...02..034M}).
The readers may turn to \citet{2012JCAP...06..036S} for detailed information about the GP method
and the package GaPP. Using the GP method, the reconstructed $H(z)$ function (solid line)
with $1\sigma$ and $2\sigma$ confidence regions (shaded areas) for 31 discrete $H(z)$ measurements
are shown in Figure~\ref{f1}(a). For comparison, we also fit the discrete $H(z)$ data using the flat
$\Lambda$CDM model. The best-fit cosmological parameters are $H_{0}=67.93^{+2.69}_{-2.61}$
km $\rm s^{-1}$ $\rm Mpc^{-1}$ and $\Omega_{m}=0.324^{+0.053}_{-0.050}$. The corresponding
best-fit theoretical curve is presented in Figure~\ref{f1}(a) with a dashed line.
It is obvious that the reconstruction of $H(z)$ is consistent with the best-fit flat $\Lambda$CDM model
within $1\sigma$ confidence region, suggesting that the GP method can provide a reliable
reconstructed function from the observed data.

Within the framework of FLRW metric, the line-of-sight comoving distance can be expressed as
\citep{1999astro.ph..5116H}
\begin{equation}
D_{C}(z)=\int^{z}_{0}\frac{dz'}{H(z')}\;.
\label{eq:dp}
\end{equation}
By integrating the reconstructed $H(z)$ function and its $1\sigma$ and $2\sigma$ error bars with respect to redshift,
we can then derive the model-independent $D_{C}(z)$ function and the corresponding $1\sigma$ and $2\sigma$
confidence regions, respectively. As shown in Figure~\ref{f1}(b), the reconstructed $D_{C}(z)$
function (solid line) is also in good agreement with that determined from the best-fit
flat $\Lambda$CDM model (dashed line).

With the reconstructed comoving distance function $D_{C}(z)$ and its $1\sigma$ uncertainty $\sigma_{D_{C}}$,
the correlated luminosity distance $D^{H}_{L}$ from the $H(z)$ data can then be calculated by
\begin{equation}\label{DL_H}
\frac{D^{H}_{L}(z)}{(1+z)} = \left\lbrace \begin{array}{lll} \frac{1}{H_{0}}\frac{1}{\sqrt{|\Omega_{K}|}}\sinh\left[\sqrt{|\Omega_{K}|}D_{C}(z)H_{0}\right]~~{\rm for}~~\Omega_{K}>0\\
                                         D_{C}(z)~~~~~~~~~~~~~~~~~~~~~~~~~~~~~~~~~~~~~~~~~~~~{\rm for}~~\Omega_{K}=0\;, \\
                                         \frac{1}{H_{0}}\frac{1}{\sqrt{|\Omega_{K}|}}\sin\left[\sqrt{|\Omega_{K}|}D_{C}(z)H_{0}\right]~~~~{\rm for}~~\Omega_{K}<0\\
\end{array} \right.
\end{equation}
with its corresponding uncertainty
\begin{equation}
\sigma_{D^{H}_{L}} = \left\lbrace \begin{array}{lll} (1+z)\cosh\left[\sqrt{|\Omega_{K}|}D_{C}(z)H_{0}\right]\sigma_{D_{C}}~~{\rm for}~~\Omega_{K}>0\\
                                         (1+z)\sigma_{D_{C}}~~~~~~~~~~~~~~~~~~~~~~~~~~~~~~~~~~~~~~~{\rm for}~~\Omega_{K}=0\;, \\
                                         (1+z)\cos\left[\sqrt{|\Omega_{K}|}D_{C}(z)H_{0}\right]\sigma_{D_{C}}~~~~{\rm for}~~\Omega_{K}<0\\
\end{array} \right.
\end{equation}
where we emphasize that the cosmic curvature $\Omega_{K}$ and the Hubble constant $H_{0}$
are the only two free parameters.

\begin{figure}
\vskip-0.2in
\centerline{\includegraphics[angle=0,width=1.15\hsize]{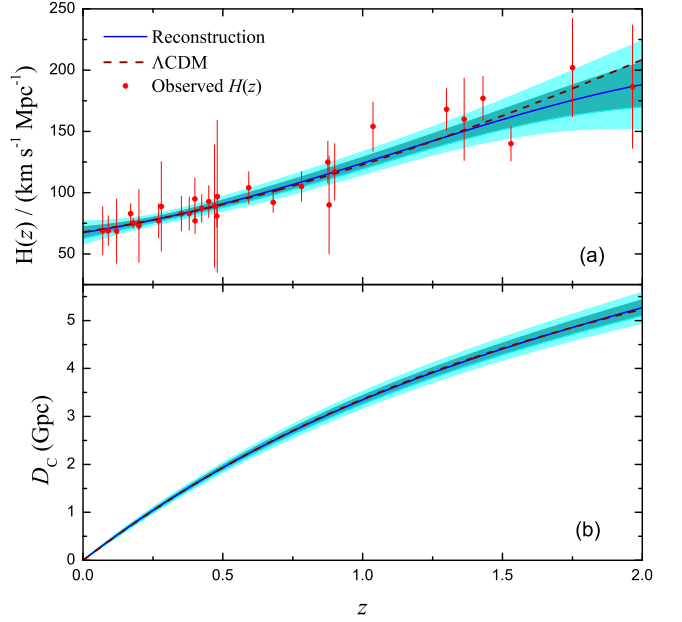}}
\vskip-0.3in
\caption{Reconstructed Hubble parameter function $H(z)$ (panel (a); solid line)
from 31 cosmic-chronometer measurements with the GP method. The corresponding continuous
$D_{C}(z)$ function (solid line) in panel (b) are derived from the reconstruction of $H(z)$.
The shadow areas are the $1\sigma$ and $2\sigma$ confidence regions of the reconstruction.
The best-fit flat $\Lambda$CDM model (dashed line) with $H_{0}=67.93$ km $\rm s^{-1}$ $\rm Mpc^{-1}$
and $\Omega_{m}=0.324$ is also shown.}
\label{f1}
\end{figure}

\subsection{Distance from GW Sources}
Unlike the distance estimation of SNe Ia that rely on the light curve fitting parameters,
the GW signals from inspiralling and merging compact binaries can provide an absolute
measure of the luminosity distance $D_{L}^{\rm GW}$. If compact binaries are NS--NS binaries or black hole
(BH)--NS binaries, the source redshifts may be obtained from EM counterparts that occur
coincidentally with the GW events \citep{2010ApJ...725..496N,2010CQGra..27u5006S,2011PhRvD..83b3005Z,2017PhRvD..95d4024C}.
Therefore, this provides a model-independent way to construct the $D_{L}^{\rm GW}$--$z$ relation.
The ET,\footnote{The Einstein Telescope Project, https://www.et.gw.eu/et/.} with ultra-high sensitivity (10 times more sensitive than the current
advanced ground-based detectors) and wide frequency range ($1-10^{4}$ Hz), would be able to
detect GW signals up to redshift $z\sim2$ for the NS--NS mergers and $z>2$ for the BH--NS systems.
Here, we forecast the curvature constraints from future GW data using the ET as reference.

To get the uncertainties in the luminosity distance of GW sources, one needs to generate the waveform
of GWs. In the transverse-traceless gauge, the detector response to a GW signal is a linear
combination of two wave polarizations,
\begin{equation}
h(t)=F_+(\theta, \phi, \psi)h_+(t)+F_\times(\theta, \phi, \psi)h_\times(t)\;,
\end{equation}
where $h_+$ and $h_\times$ are the plus and cross modes of GW, respectively.
The antenna pattern functions $F_{+}$ and $F_\times$ depend on the source's position ($\theta, \phi$),
the polarization angle $\psi$, as well as the detector's location and orientation.
The pattern functions of one of the three interferometers in the ET are~\citep{2011PhRvD..83b3005Z}
\begin{align}
F_+^{(1)}(\theta, \phi, \psi)=&~~\frac{{\sqrt 3 }}{2}[\frac{1}{2}(1 + {\cos ^2}(\theta ))\cos (2\phi )\cos (2\psi ) \nonumber\\
                              &~~- \cos (\theta )\sin (2\phi )\sin (2\psi )],\nonumber\\
F_\times^{(1)}(\theta, \phi, \psi)=&~~\frac{{\sqrt 3 }}{2}[\frac{1}{2}(1 + {\cos ^2}(\theta ))\cos (2\phi )\sin (2\psi ) \nonumber\\
                              &~~+ \cos (\theta )\sin (2\phi )\cos (2\psi )].
\label{equa:F}
\end{align}
Since these interferometers align with an angle $60^\circ$ with each other,
the two others' pattern functions are $F_{+,\times}^{(2)}(\theta, \phi, \psi)=F_{+,\times}^{(1)}(\theta, \phi+2\pi/3, \psi)$
and $F_{+,\times}^{(3)}(\theta, \phi, \psi)=F_{+,\times}^{(1)}(\theta, \phi+4\pi/3, \psi)$, respectively.

Following \citet{2009LRR....12....2S} and \citet{2011PhRvD..83b3005Z}, we calculate the Fourier transform $\mathcal{H}(f)$
of the time domain waveform $h(t)$ by applying the stationary phase approximation,
\begin{equation}
\mathcal{H}(f)=\mathcal{A}f^{-7/6}e^{i\Psi(f)}\;,
\label{equa:hf}
\end{equation}
where the Fourier amplitude is given by
\begin{align}
\mathcal{A}=&~~\frac{1}{D_{L}^{\rm GW}}\sqrt{F_+^2\left(1+\cos^2(\iota)\right)^2+4F_\times^2\cos^2(\iota)}\nonumber\\
            &~~\times \sqrt{5\pi/96}\pi^{-7/6}\mathcal{M}_c^{5/6}\;,
\label{equa:A}
\end{align}
where $\iota$ is the inclination angle between the binary's orbital and the line-of-sight, and
\begin{equation}
D_{L}(z)=\frac{1+z}{H_0}\int_{0}^{z}\frac{{\rm d}z}{\sqrt{\Omega_{m} (1+z)^{3}+1-\Omega_{m}}}
\label{equa:DL}
\end{equation}
is the theoretical luminosity distance in the flat $\Lambda$CDM model.
Here $\mathcal{M}_{c}=(1+z)M \eta^{3/5}$ represents the observed chirp mass, where
$M=m_1+m_2$ is the total mass of binary components, and $\eta=m_1 m_2/M^2$ denotes the symmetric mass ratio.
The expression of the function $\Psi$ can be found in~\citet{2011PhRvD..83b3005Z}.
Averaging the Fisher matrix over the inclination $\iota$ and the polarization $\psi$
with the limit $\iota<20^\circ$ is nearly equivalent to taking $\iota=0$.
In the following simulations, we take the simplified case of $\iota=0$, as
\citet{2017PhRvD..95d4024C} did in their treatment.

The combined signal-to-noise ratio (SNR) for the network of three independent ET interferometers
is given by
\begin{equation}
\rho=\sqrt{\sum\limits_{i=1}^{3}\left\langle \mathcal{H}^{(i)},\mathcal{H}^{(i)}\right\rangle}\;,
\label{euqa:rho}
\end{equation}
where the inner product is defined as
\begin{equation}
\left\langle{a,b}\right\rangle=4\int_{f_{\rm lower}}^{f_{\rm upper}}\frac{\tilde a(f)\tilde b^\ast(f)+\tilde a^\ast(f)\tilde b(f)}{2}\frac{{\rm d}f}{S_h(f)}\;,
\label{euqa:product}
\end{equation}
where $\thicksim$ represents the Fourier transformation, $S_h(f)$ is the one-side noise power spectral density
characterizing the performance of the GW detector, $f_{\rm lower}$ and $f_{\rm upper}$
are the lower and upper cutoff frequencies. Here we adopt $f_{\rm lower}=1$ Hz and $f_{\rm upper}=2f_{\rm LSO}$,
where the orbit frequency at the last stable orbit $f_{\rm LSO}=1/(6^{3/2}2\pi M_{\rm obs})$ with the observed
total mass $M_{\rm obs}=(1+z)M$~\citep{2011PhRvD..83b3005Z}.
The signal is claimed as a GW event only when the SNR of the detector network reaches over 8 (i.e., $\rho >8.0$).

The instrumental uncertainty on the measurement of $D_{L}^{\rm GW}$ can be estimated by using the Fisher matrix.
Assuming that the uncertainty of $D_{L}^{\rm GW}$ is irrelevant to the uncertainties of the remaining GW parameters,
we have~\citep{2011PhRvD..83b3005Z}
\begin{align}
\sigma_{D_L}^{\rm inst}\simeq \sqrt{\left\langle\frac{\partial \mathcal H}{\partial D_{L}^{\rm GW}},\frac{\partial \mathcal H}{\partial D_{L}^{\rm GW}}\right\rangle^{-1}}\;.
\end{align}
As $\mathcal H \propto 1/D_L^{\rm GW}$, we can derive $\sigma_{D_L}^{\rm inst}\simeq D_L^{\rm GW}/\rho$.
Considering the maximal effect of the inclination $\iota$ on the SNR, we add a factor of 2 to the instrumental uncertainty
for a conservative estimation
\begin{align}
\sigma_{D_L}^{\rm inst}\simeq \frac{2D_L^{\rm GW}}{\rho}\;.
\label{sigmainst}
\end{align}
We also add an additional error $\sigma_{D_L}^{\rm lens}/D_L^{\rm GW}=0.05z$ due to the weak lensing effect.
Thus, the total uncertainty on $D_{L}^{\rm GW}$ is taken to be
\begin{equation}
\sigma_{D_L^{\rm GW}}=\sqrt{\left(\frac{2D_L^{\rm GW}}{\rho}\right)^2+\left(0.05z D_L^{\rm GW}\right)^2}\;.
\label{eq:sigmaDL}
\end{equation}

\begin{figure*}
\vskip-0.1in
\centerline{\includegraphics[keepaspectratio,clip,width=1.0\textwidth]{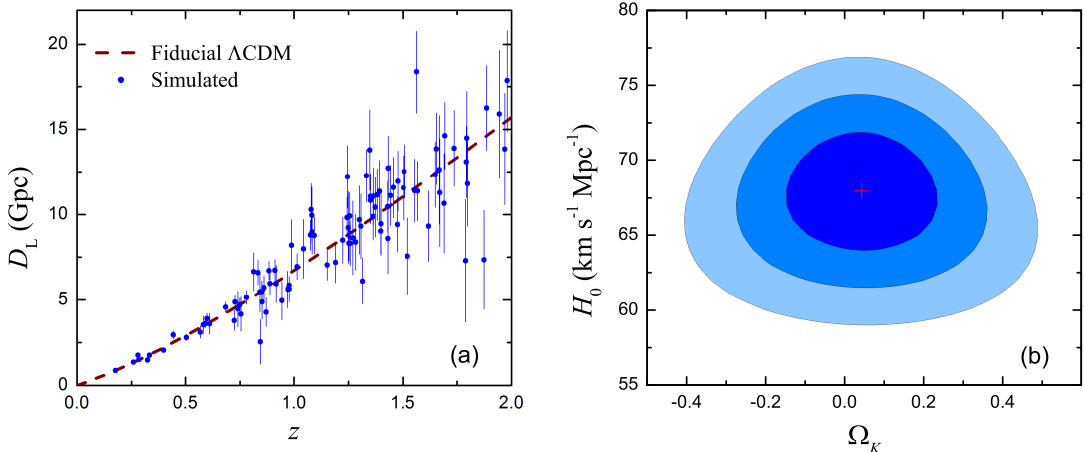}}
\vskip-0.1in
\caption{Panel (a): an example catalogue of 100 simulated GW events with redshifts $z$ and luminosity distances $D_{L}$.
The dashed line denotes the fiducial $\Lambda$CDM model.
Panel (b): $1-3\sigma$ constraint contours for $\Omega_{K}$ and $H_{0}$, using 100 simulated GW events.
The cross represents the best-fit pair.}
\label{f2}
\end{figure*}

\section{Monte Carlo Simulations}
\label{sec:simulation}
In this section, we perform Monte Carlo simulations to test how well GW standard sirens and cosmic chronometers can be used
to constrain the cosmic curvature. To do so, we have to choose a fiducial cosmological model. Note that the exact value of the curvature
parameter will not be essential in our simulations, since we are only interested in the precision with which it can be constrained.
However, for consistency with current expansion rate measurements, we adopt their best-fit cosmological parameters
in the fiducial flat $\Lambda$CDM model: $H_{0}=67.93$ km $\rm s^{-1}$ $\rm Mpc^{-1}$, $\Omega_{m}=0.324$,
and $\Omega_{\Lambda}=1-\Omega_{m}$. Following the process of producing the mock GW data in \citet{2017PhRvD..95d4024C},
we expect the source redshift can be obtained by identifying an EM counterpart of the GW event, and simulate
many catalogues of NS--NS and BH--NS systems with their $z$, $D_{L}^{\rm GW}$, and $\sigma_{D_L}^{\rm GW}$.
The redshift distribution of the observable sources is adopted as
\citep{2011PhRvD..83b3005Z}
\begin{equation}
P(z)\propto \frac{4\pi D_C^2(z)R(z)}{H(z)(1+z)},
\label{equa:pz}
\end{equation}
where $D_C(z)$ is the comoving distance, and $R(z)$ describes the time evolution of the burst rate, which is given by
\citep{2001MNRAS.324..797S,2009PhRvD..80j4009C}
\begin{equation}
R(z)=\begin{cases}
1+2z, & z\leq 1 \\
\frac{3}{4}(5-z), & 1<z<5 \\
0, & z\geq 5.
\end{cases}
\label{equa:rz}
\end{equation}
We sample the source redshift $z$ from the probability distribution function (Equation~(\ref{equa:pz})).
To be consistent with the redshift range of current expansion rate measurements, we consider the potential
observations of GW standard sirens in $0< z < 2.0$. With the mock $z$, we calculate the fiducial value
of $D_{L}^{\rm fid}$ based on Equation~(\ref{equa:DL}). The masses of each NS and BH are chosen to be uniform
in the intervals $[1,2]$ and $[3,10]$ $M_{\odot}$, respectively. The ratio of possibly detecting BH--NS
and NS--NS events is assumed to be 0.03, as predicted for the Advanced LIGO-Virgo network \citep{2010NIMPA.624..223A}.
The position angle $\theta$ is randomly sampled within the interval $[0,\pi].$\footnote{Because the SNR is
independent of the other two angles $\phi$ and $\psi$, we do not need to consider them.}
We then evaluate the combined SNR for each set of the random values using Equation~(\ref{euqa:rho}),
and confirm that it is a successful GW event detection if $\rho >8.0$. For every confirmed detection,
we add the deviation $\sigma_{D_{L}^{\rm GW}}$ in Equation~(\ref{eq:sigmaDL}) to the fiducial value of $D_{L}^{\rm fid}$.
That is, we sample the $D_{L}^{\rm GW}$ measurement according to the Gaussian distribution
$D_{L}^{\rm GW}=\mathcal{N}(D_{L}^{\rm fid},\;\sigma_{D_{L}^{\rm GW}})$.
The detection rates of BH--NS and NS--NS per year for the ET are estimated to be about the order $10^{3}-10^{7}$ $\rm events$ $\rm yr^{-1}$.
However, only a small fraction ($\sim 10^{-3}$) is predicted to have the observation of EM counterpart.
Taking the detection rate in the middle rang $\mathcal{O}(10^5)$, and assuming that the fraction of
the observation of EM counterpart is the same at any time interval, we can expect to detect $\mathcal{O}(10^2)$
GWs with EM counterparts per year. Note that we are only interested in what level of curvature constraints
can be achieved by a certain number of future GW data together with their EM counterparts providing source redshifts.
The use of the GW detection rate and the fraction of the observation of EM counterpart are therefore not essential
in our simulations. We first simulate a population of 100 GW events with redshifts $z$,
luminosity distances $D_{L}^{\rm GW}$, and the errors of the luminosity distances $\sigma_{D^{\rm GW}_{L}}$.

An example of 100 simulated GW events from the fiducial model is presented in Figure~\ref{f2}(a).
By confronting distances $D_{L}^{\rm GW}(z)$ from the simulated GW events with distances $D_{L}^{H}(z)$
in Equation~(\ref{DL_H}) that depend on $\Omega_{K}$ and $H_0$ from observations of cosmic chronometers,
we can obtain a model-independent estimation for the cosmic curvature by minimizing the $\chi^{2}$ statistic:
\begin{equation}
\chi^2(\Omega_{K},\;H_0)=\sum_{i}\frac{\left[D_{L}^{H}(z_{i};\;\Omega_{K},\;H_0)-D_{L}^{\rm GW}(z_{i})\right]^{2}}
{\sigma_{D^{H}_{L},i}^{2}+\sigma_{D^{\rm GW}_{L},i}^{2}}\;.
\end{equation}
To ensure the final constraints are unbiased, we repeat the simulation process 1000 times for
each data set by using different noise seeds. Figure~\ref{f2}(b) shows the constraint results on
$\Omega_{K}$ and $H_0$.\footnote{Here we use a Gaussian prior of $H_0=67.93\pm2.60$ km
$\rm s^{-1}$ $\rm Mpc^{-1}$ to guide the minimization procedure over $H_0$.}
We find that, from 100 simulated GW events and observations of cosmic chronometers, the model-independent
estimation for the cosmic curvature is $\Omega_{K}=0.043^{+0.125}_{-0.124}$ ($1\sigma$).
At this point, it is interesting to compare our forecast result with some actual model-independent tests involving the
distances from other popular astrophysical probes. The error on the determined $\Omega_{K}$
is at the level of $\sigma_{\Omega_{K}}\simeq0.125$ with 100 GW events, which is 40\% smaller than
that of the Union2.1/JLA SNe Ia ($\sigma_{\Omega_{K}}\simeq0.20$;~\citealt{2016ApJ...833..240L,2017ApJ...838..160W}),
and is 60\% smaller than that of 120 radio quasars ($\sigma_{\Omega_{K}}\simeq0.29$;~\citealt{2017arXiv170808608C}).
Therefore, we can conclude that in the framework of model-independent methods testing the spatial curvature,
GWs may achieve constraints with higher precision.

To better represent how effective our method might be with a certain number of GW events,
in Figure~\ref{f3} and Table~\ref{table2} we display the best-fit $\Omega_{K}$ and $1\sigma$ confidence level
as a function of the number of GW events $N$. The model-independent test of $\Omega_{K}$
from 580 Union2.1 SNe Ia (black diamond; \citealt{2016ApJ...833..240L}) is also plotted for comparison.
One can see from Figure~\ref{f3} that as the number of GW events increases, the uncertainty of $\Omega_{K}$
is reduced. The precision of the determined $\Omega_{K}$ from 100 GW events is already better than
that of 580 Union2.1 SNe Ia. The spatial curvature can be constrained with an error of only 0.04 if 1000 GW events are observed.

\begin{figure}
\vskip-0.1in
\centerline{\includegraphics[keepaspectratio,clip,width=0.55\textwidth]{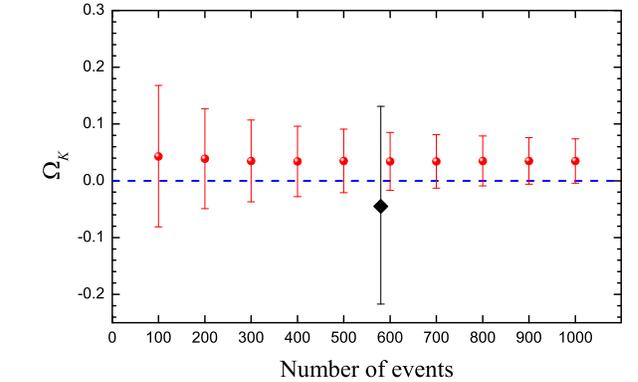}}
\vskip-0.1in
\caption{Best-fit $\Omega_{K}$ and $1\sigma$ confidence level as a function of the number of GW events.
The black diamond represents the model-independent constraint from 580 Union2.1 SNe Ia. The blue dashed line is the fiducial value.}
\label{f3}
\end{figure}

\begin{table}
\centering \caption{Summary of Model-independent Curvature Determinations from
$N$ Simulated GW Events and Observations of Cosmic Chronometers}
\begin{tabular}{cc|cc}
\hline
\hline
 $N^{a}$ &  $\Omega_{K}$  &  $N^{a}$  &  $\Omega_{K}$ \\
\hline
100   &   $0.043^{+0.125}_{-0.124}$   &   600   &   $0.034\pm0.051$   \\
200   &   $0.039\pm0.088$   &   700   &   $0.034\pm0.047$   \\
300   &   $0.035\pm0.072$   &   800   &   $0.035\pm0.044$   \\
400   &   $0.034\pm0.062$   &   900   &   $0.035\pm0.041$   \\
500   &   $0.035\pm0.056$   &   1000   &   $0.035\pm0.039$   \\
\hline
\end{tabular}
\label{table2}
\medskip \\
$^{\rm a}$$N$ denotes the number of GW events.
\end{table}

By the time we have ET results, there might be other $H(z)$ measurements with
wider redshift range and higher accuracy from different observables.
To investigate the case of adding more cosmic-chronometer measurements, we also
perform Monte Carlo simulations to create the mock $H(z)-z$ data sets. We assume that
there are other 50 mock $H(z)-z$ data points by the time that 1000 GW events are detected,
the redshifts of which are chosen equally in log$(1+z)$ space in $0.1\leq z \leq5.0$.
The relative uncertainty of these mock data is taken at a level of 1\%, which will be
realized in future observations \citep{2013PhR...530...87W}. The route of GW simulation
is the same as described earlier in Section~\ref{sec:simulation}, but now we consider
the potential observations of GW standard sirens in $0 < z < 5.0$.
Figure~\ref{f4} gives an example of the simulations in the case of adding 50 cosmic-chronometer measurements.
From top to bottom, the three panels show the cosmic-chronometer data (including 31 observed $H(z)$
data (solid points) and 50 mock $H(z)$ data (circles)) with the reconstructed $H(z)$ function (solid line),
the reconstructed $D_{C}(1+z)$ function (solid line) and 1000 simulated GW events with luminosity distances $D_{L}$ (solid points), and
the final constraints on $\Omega_{K}$ and $H_0$, respectively. In this case, the final derived $\Omega_{K}$
is $\Omega_{K}=-0.002\pm0.028$ ($1\sigma$). Compared with the constraint obtained from 1000 simulated GW events and 31 current
cosmic-chronometer measurements ($\Omega_{K}=0.035\pm0.039$), the uncertainty of the determined $\Omega_{K}$
in this case can be further improved by a factor of 1.4.

\begin{figure}
\vskip-0.3in
\centerline{\includegraphics[keepaspectratio,clip,width=0.45\textwidth]{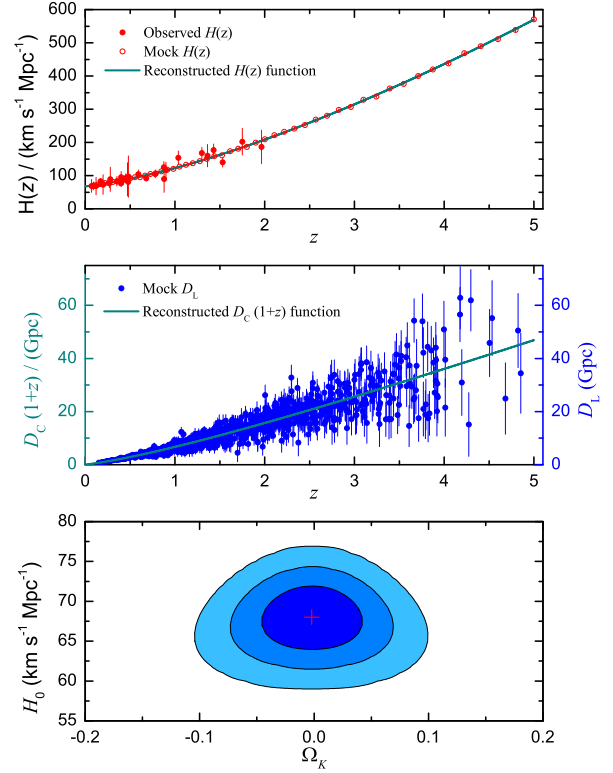}}
\vskip-0.4in
\caption{An example of the simulations for the case of 81 cosmic-chronometer measurements
and 1000 simulated GW events.
Top panel shows the cosmic-chronometer data (including 31 observed $H(z)$ data (solid points)
and 50 mock $H(z)$ data (circles)) with the reconstructed $H(z)$ function (solid line).
Middle panel shows the reconstructed $D_{C}(1+z)$ function (solid line) and 1000 mock $D_{L}$ data (solid points).
Bottom panel shows the final constraints on $\Omega_{K}$ and $H_0$ from these data.}
\label{f4}
\end{figure}

\section{Summary and discussion}
\label{sec:summary}
The coincident detection of gravitational and EM waves from a binary NS merger has formally opened a new window
on observational cosmology. More precisely, the greatest advantage of GW standard sirens is that
the distance calibration is independent of any other distance ladders.
In this work, we investigate the constraint ability of future GW observations of the ET on the spatial curvature
by using a model-independent method. The main principle of our method is to compare two kinds of luminosity distances.
One distance $D_{L}^{H}(\Omega_{K},\;H_0)$ is constructed with Hubble parameter measurements obtained from observations
of cosmic chronometers, which is susceptible to the curvature parameter $\Omega_{K}$ and the Hubble constant $H_0$.
Based on the discrete Hubble parameter data, we first use the GP method to reconstruct the continuous $H(z)$ function. Next,
we obtain the model-independent comoving distance function $D_{C}(z)$ by directly calculating the integral of the reconstructed $H(z)$ function.
Using this continuous $D_{C}(z)$ function, the luminosity distance $D_{L}^{H}(\Omega_{K},\;H_0)$ from the $H(z)$
measurements can be further calculated at a certain $z$. The other distance $D_{L}^{\rm GW}$ is from the simulated GW data,
which is independently determined. Previously, by confronting $D_{L}^{H}(\Omega_{K},\;H_0)$ with luminosity distances
from observations of SNe Ia, some studies achieved model-independent constraints on the spatial curvature
\citep{2016ApJ...833..240L,2017ApJ...838..160W}. However, the constraint ability of SNe Ia are obviously restricted
by the fact that their distances depend on light-curve fitting parameters. While GW standard sirens have
the advantage of being self-calibrating. Therefore, combining the GW observations with $H(z)$ data may provide
a powerful and novel way to estimate the spatial curvature.

Through Monte Carlo simulations, we find that the error of the curvature parameter can be expected to be constrained
at the level of $\sim0.125$ by combining 31 current observed $H(z)$ data and 100 simulated GW data. The uncertainty of $\Omega_{K}$ can be further reduced to
$\sim0.04$ if 1000 GW events are observed. We also find that with 81 cosmic-chronometer measurements
(including 31 observed $H(z)$ data and 50 mock $H(z)$ data) and 1000 simulated GW events, one can expect the zero
cosmic curvature to be estimated at the precision of $\Omega_{K}=-0.002\pm0.028$. By comparing our results with previous ones which reported model-independent
curvature tests using current data of SNe Ia and radio quasars \citep{2016ApJ...833..240L,2017arXiv170808608C,2017ApJ...838..160W},
we demonstrate that future measurements of the luminosity distances of GW sources will be more competitive than current analyses.
These results show that the prospects for testing the spatial curvature with GW observations is very promising.

\acknowledgments
We are grateful to the anonymous referee for constructive suggestions.
This work is partially supported by the National Basic Research Program (``973'' Program)
of China (grant No. 2014CB845800), the National Natural Science Foundation of China
(grant Nos. U1831122, 11603076, 11673068, and 11725314), the Youth Innovation Promotion
Association (2011231 and 2017366), the Key Research Program of Frontier Sciences (grant No. QYZDB-SSW-SYS005),
the Strategic Priority Research Program ``Multi-waveband gravitational wave Universe''
(grant No. XDB23000000) of the Chinese Academy of Sciences, and the Natural Science Foundation
of Jiangsu Province (grant No. BK20161096).

\bibliographystyle{apj}

\end{document}